\input psfig.sty
\documentstyle{l-aa}     % LaTeX A&A  Standard Fonts

\def\sun{\hbox{$\odot$}}

\begin{document}

   \thesaurus{11.01.2;11.07.1; 11.19.1    \qquad      A\&A Section : 03
             }

   \title{BLR sizes and the X-ray spectrum in AGN}

   \author{A. Wandel\inst{1}, Th. Boller\inst{2}
                     }

   \offprints{ A. Wandel
             }

   \institute{$^1$ Racah Institute of Physics, The Hebrew University of
                    Jerusalem 91904, Israel\\
$^2$ Max-Planck-Institut f\"ur Extraterrestrische Physik,
             Postfach 1603, D-85740 Garching, Germany\\
              }

   \date{Received 18 November 1996 / Accepted 20 October 1997   }
%   \date{Received\ \ \ ; accepted \ \ \  }

   \maketitle

   \begin{abstract}
Recent ROSAT studies of narrow-line Seyfert 1 galaxies
revealed these objects
(with FWHM $H_{\beta} \le 2000\ km\ s^{-1}$)
generally show steeper soft X-ray spectra than
broad line Seyfert 1's and that there are no AGN with broad lines
and steep X-ray spectra.
We derive a simple theoretical model which explains this observed correlation
between the line width and the spectral index for Seyfert 1 galaxies.
Assuming the line width is due to gravitational velocity dispersion,
it  is determined by the radius of the broad line region.
Sources with steep X-ray spectra (for a given luminosity)
have a stronger ionizing power than flat-spectrum sources with comparable
luminosity, which implies
that the BLR is formed at relatively larger distances from
the central source, and hence has a smaller velocity dispersion
and a smaller observed FWHM.
We test the model over a hetrogeneous (normal and narrow-line) sample of 
some 50 AGN finding a good agreement with the data.


\end{abstract}

\section{Introduction}

Boller, Brand \& Fink 1996 (hereafter BBF96) report the observation
with ROSAT of a sample of 49 Narrow-Line Seyfert 1 galaxies (NLS1), finding
that NLS1 have generally steeper soft X-ray continua than normal
Seyfert 1s. (However, not all NLS1 are remarkably steep).
As a result, looking in the FWHM-$\alpha$ plane,
there are no AGN with broad lines and steep X-ray spectra.

BBF96 tested  NLS1 models assuming they are Seyfert 1 galaxies with 
 pole-on orientation,  warm absorption, thick BLR and
smaller black hole mass and/or higher accretion rates.
The latter showed some promise, but have drawbacks as well,
as some NLS1s do not have steep spectra, and on the other hand
a smaller black hole mass does not necessarily produce narrower lines
(see Sect. 2).

We suggest a simple physical explanation to the observed correlation
between the soft X-ray spectral index and the H$\beta$ line width.
Assuming the narrower permitted lines  reflects a lower Keplerian velocity, we
 explain the observed correlation 
 by showing that a steeper spectrum has a stronger
ionizing power, and hence the BLR
form at a relatively larger distance from the central source. 
(Indeed, the characteristic distance of the BLR
estimated by this method, agrees well with the BLR sizes
determined by reverberation mapping (Wandel 1996;1997)).

Below we formulate this
scanario by combining the assumptions of
(1) Keplerian velocity Doppler width for
the broad lines, (2) a power-law
spectrum of the ionizing continuum luminosity, and
(3) characteristic values for the ionization parameter and physical conditions
in the BLR.

\section{The analytic approach}

%> Fig. 1 %
\begin{figure}
       \psfig{figure=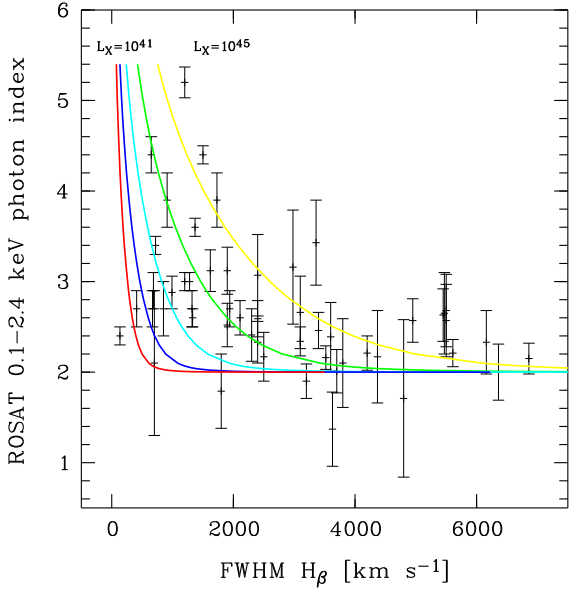,height=8.7cm,clip=}
      \caption{X-ray continuum slope, obtained from a power-law fit,
               versus FWHM of H$\beta$ for Seyfert 1 galaxies (taken
               from Walter \& Fink 1993)  and NLS1 (BBF96).
               Lines of constant X-ray luminosity are overplotted for the
               relation given by Eq. 7.}
\end{figure}

We derive a simple analytic relation between the width of broad
emission lines and the continuum slope of the central engine.
The observed velocity dispersion is determined by the mass of the
black hole and the Broad line region radius.
%*low mass black holes will produce low Keplerian velocities;
%> This is wrong!
(Note however that the BH mass does not solely determine the BLR velocity;
for example, in Bondi accretion we have $\dot M \sim R_a^2$ where 
$R_a\sim GM/kT$ is the accretion radius, so that $\dot M\sim M^2$. If the
efficiency is constant, $L\sim M^2$
 than $R\sim L^{\frac{1}{2}}\sim M$ so
 $v\propto M/R\sim const$).
Our basic idea is as follows:
Sources with steep X-ray spectra (for a given luminosity)
have a stronger ionizing power than sources with flat spectra.
%* in conjunction with the shift of the peak emissivity...
%> the location of the peak emissivity is irrelevant, what determines
%> the FWHM is the velocity _dispersion_.
Since the ionization parameter at the BLR
has a characteristic value, a larger ionizing power implies that
the BLR is formed at a relatively larger distance from
the central source, and hence has a smaller velocity dispersion and
a smaller observed FWHM.

%> Fig. 2 $
\begin{figure}
       \psfig{figure=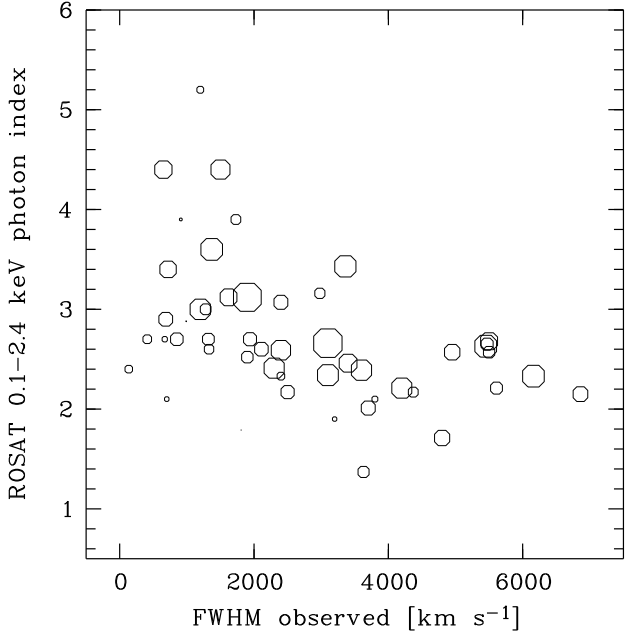,height=8.7cm,clip=}
\caption{X-ray continuum slope, obtained from a power-law fit,
versus FWHM of H$\beta$ for Seyfert 1 galaxies (taken
from Walter \& Fink 1993)
and NLS1 (BBF96).The size of the circles corresponds to
the ROSAT luminosity.
}
\end{figure}

We make the following assumptions (rather standard in the literature) :

a. The width of the broad lines is induced by  Keplerian motion in the
  gravitational potential of the central mass.
  The full width at half maximum  is given by:

$$ FWHM \approx \left(\frac{GM}{R}\right)^{\frac {1}{2}}    \eqno(1)  $$
where M is the mass of the central black hole
and R the radius of the broad line region.

b. The physical conditions in the ionized gas emitting the broad lines are
characterized by the ionization parameter U, the ratio of ionizing photons to
electrons (cf. Netzer 1990) defined by

$U = Q_{ion} /{4\pi R^2} c n_e$
%> I have removed f_c in the above eqn since there is NO covening factor
%>in the definition of U!
where
$Q_{ion}= \int_{E_0}^\infty l(E) \frac{dE}{E}$
is the ionizing photon flux (number of ionizing photons per unit time)
l(E) is the monochromatic luminosity of the central source, per unit
energy, and $n_e$ is the electron density.

The radius of the BLR may be written as
$$
R= \left( \frac{L_{ion}}{ 4\pi c\bar E_{ion} U n_e}\right ) ^{1/2}
%= 0.04 \left ( \frac {f_c L_{45}}{U n_{10} \bar E} \right )^{1/2} pc
\eqno (2)
$$
where
$L_{ion} = \int_{E_0}^\infty l(E) dE$ is the ionizing luminosity,
and $\bar E_{ion} \equiv L_{ion}/Q_{ion}$,
%* and $\bar E_{ion} = \frac{L_{ion}}{/Q_{ion} 4\pi R^2} $,
is the mean energy of the  ionizing photons.

Analyses of the broad emission lines in various AGN indicates that
typical values of U for high ionization lines such as H$\beta$
in AGN clouds are
0.1-1 and
$n_e \sim 10^{10}-10^{11} cm^{-3}$
in the high excitation lines (cf. Rees, Netzer \& Ferland 1989)
so that $Un \sim 10^{9-11} cm^{-3}$.
%For $nU = 10^{10}$ eq. (2) gives
$$
R \simeq 0.037 (n_{10}U\bar E_{ion})^{-1/2}  L_{i45}^{1/2} pc \eqno (3)
$$
where $L_{i45}=L_{ion}/10^{45}$erg/s,
$n_{10}=n_e/10^{10}cm^{-3}$
and $\bar E_{ion}$ is in Rydbergs.

This value (cf. Alexander \& Netzer 1994), is in agreement with the
results from recent reverberation mapping observations
(e.g. Clavel et al. 1991, Peterson et al. 1991)

c. Assuming the energy spectrum of the luminosity emerging from the
central energy source can be
approximated by a power law $l(E) \propto E^{-\alpha}$ we find

$$
\bar E_{ion} = \frac {\alpha}{\alpha-1}E_0. \ \ \ \alpha > 1\ \eqno (4)
$$

(Since we are interested only in the ionizing
spectrum, between $E_0=$1 Rydberg and a few keV, the power law assumption
may be a reasonable approximation.
Equation 4 holds only for $\alpha > 1$, as otherwise $L_{ion}$ and E diverge.
This divergence is not physical, as actually the hard X-ray spectrum cuts off at
at 50-100 keV).

Combining Eqs. (1), (3) and (4)  we have

$$
FWHM \approx (1100 km/s)
     \left(\frac{n_{10}U\alpha}{\alpha-1}\right)^{1/4} L_{i45}^{-1/4}
     M_7^{1/2}
\eqno (5)
$$
where $M_7=M/10^7M_\odot$.
Since we do not know the ionizing luminosity, we try to express it in terms of
the X-ray luminosity and the spectral slope.
The observed X-ray luminosity $L_x$ is in the ROSAT 0.1-2.4 keV band,
while the ionizing luminosity extends to 1Ryd=13.6eV. Assuming the power law spectrum in
the ROSAT band may be extrapolated to a lower energy $E_0$ we have for the ratio

$${L_{ion}\over
L_x}=\frac{E_0(keV)^{1-\alpha}}{0.1^{1-\alpha}-2.4^{1-\alpha}} \approx
(0.14 E_0)^{1-\alpha }    \eqno (6),$$ 
where here and below $E_0$ without units indicates $E_0$ in Rydbergs.
Substituting this in Eq. (5) we have
$$
FWHM \approx (1100 km/s)
     \left(\frac{n_{10}U\alpha}{\alpha-1}\right)^{1/4} L_{x45}^{-1/4}  
$$
$$
    (0.14 E_0)^{(\alpha -1)/4} M_7^{1/2}
\eqno (7)
$$
where $L_{x45}=L_x/10^{45}$erg/s.

d. We do not know the BH mass of individual objects, but we can use
a mass-luminosity relation in a statistical sense, over a large enough
sample of AGN, as we did with the ionization parameter above.
We assume that the mass is roughly proportional to the luminosity,
that is,
that the Eddington ratio has  a relatively
narrow distribution, compared to the range of the luminosity distribution
over the sample (e.g. Wandel and Yahil 1985).
Defining , $\eta_x= L_x/L_{Edd}$
where
$ L_{Edd}=(1.3\ 10^{45} erg/s ) M_7$,
we can then express the central mass in terms of the continuum
X-ray luminosity,
$M_7\approx  L_{x45}/\eta_x  $.
Substitutinging this relation in Eq. (6) we have
$$
FWHM \approx (1100 km/s) \eta_x ^{-1/2}
     \left(\frac{n_{10}U\alpha}{\alpha-1}\right)^{1/4} L_{x45}^{1/4}
$$
$$
  (0.14 E_0)^{(\alpha -1)/4}
\eqno (8)
$$

Detemining $E_0$ and $\eta_x$ will give us a relation between the line width,
the spectral index and the X-ray luminosity.

The ratio $L_x/L_{Edd}$  is not
known, but may be estimated from theoretical and observational arguments.
The bolometic $L/L_{Edd}$ is often estimated to be in the range 0.1-1.
Observational estimates of $L_{opt}/L_{Edd}$
and $L_x/L_{Edd}$ are in the range 0.01-0.1 (e.g. Wandel and Yahil 1985,
Wandel \& Mushotzky 1986), which may justify a choice of $\eta_x=0.1$.

%Taking $E_0=1Ryd$ and  $\eta_x=0.1$ in 
Combining  the parameters in Eq. (8) as
 $P=(n_{10}U)^{1/4} (\eta_x/0.1)^{-1/2} $
gives
$$
FWHM \approx (5600 km/s) 1.6^{-\alpha}
  \left(\frac{E_0^{\alpha -1}\alpha}{\alpha-1}\right)^{1/4} L_{x45}^{1/4} P
\eqno (9)
$$
Note that choosing $E_0=1$  
is equivalent to extrapolating the power law spectrum to 1Ryd, 
which may overestimate the ionizing luminosity,
(especially for large values of $\alpha$) if the spectrum 
actually flattens at  a higher energy. 
%We therefore calculate the model also for a larger value of $E_0=3Ryd$.
However, because of the weak dependence on $E_0$, this effect is small:
for example, if $L_x=10^{44} erg/s$, $p=1$ and $\alpha = 2$
$E_0=1Ryd$ gives
$FWHM \simeq 1500 km/s$, while for $E_0=3Ryd$ we get 1800 km/s.

%$$FWHM \approx (4400 km/sec)
% \left(\frac{\alpha}{\alpha-1}\right)^{1/4} 1.26^{-\alpha} L_{x45}^{1/4}
%\eqno (10)$$

In Fig. 1 we overlay the analytic relation between the FWHM and $\alpha$ with
the observed data from Fig. (8) of BBF96 for different values of $L_x$.
The general shape of the observed distribution and the analytic relation of
Eq. (9) are in good agreement.
(Note that the convergence of the theoretical curves at a photon index of 2
(corresponding to $\alpha =1$) for large FWHM is not physical; if a high
energy cutoff would have been introduced, the curves would extend also to
lower values).

\section{Predictions and tests of the model}

\subsection{Line width}

%> Figs. 3ab %
\begin{figure*}
  \begin{minipage}{8.7cm}
       \psfig{figure=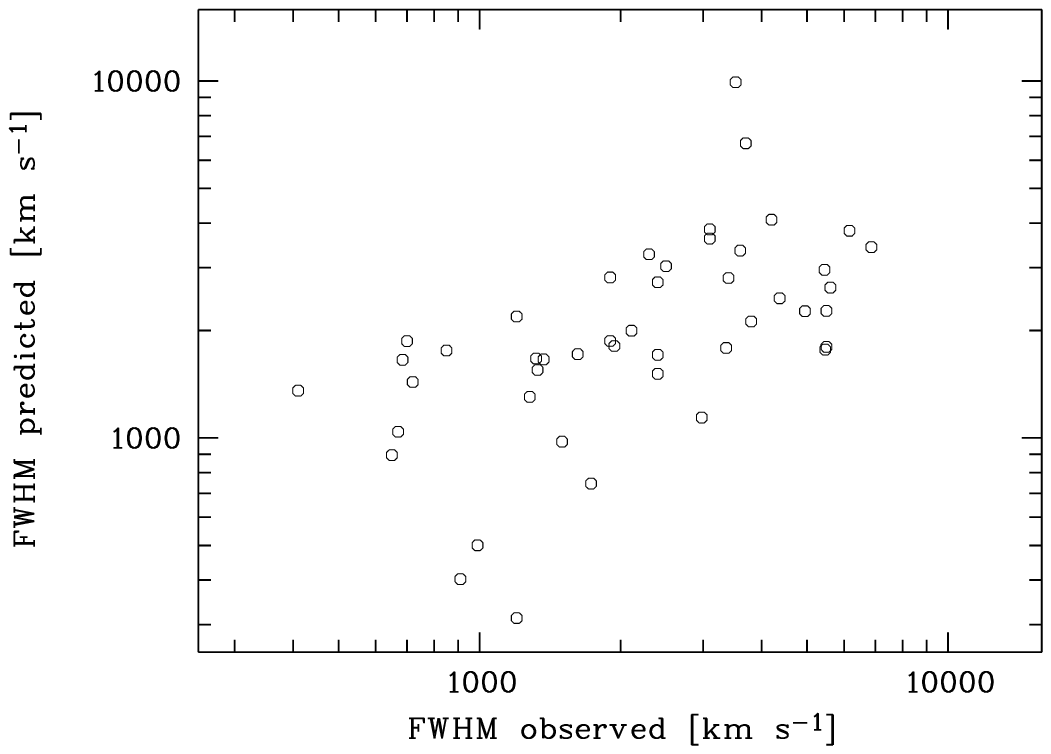,height=8.7cm,width=8.7cm,clip=}
\end{minipage}
  \begin{minipage}{8.7cm}
       \psfig{figure=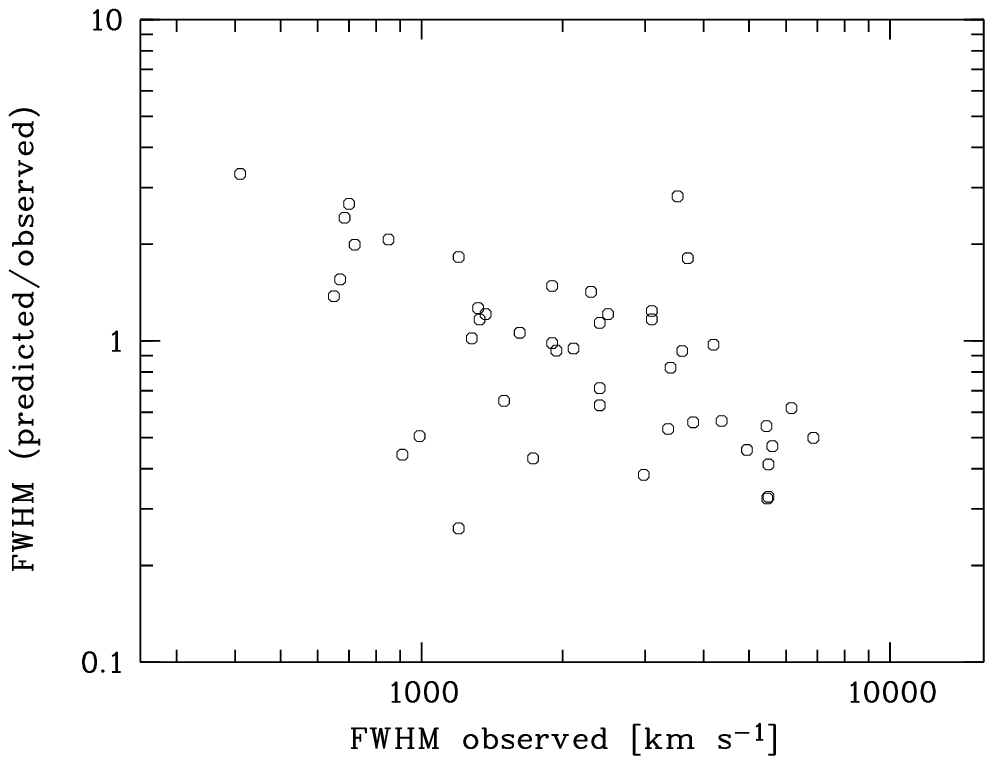,height=8.7cm,width=8.7cm,clip=}
\end{minipage}
\caption{Line width prediction from the model.
Left panel (3a):
Predicted (from Eq. 9) versus observed FWHM of $H\beta$ for Seyfert 1 galaxies
(taken from Walter \& Fink 1993) and NLS1  (BBF96).
Right panel (3b):
Ratio of predicted and observed FWHM  versus observed FWHM of $H\beta$.
}
\end{figure*}

Using Eq. (8) or (9) we may predict the line width from the observed
luminosity
and the X-ray continuum slope. Figure 3a shows the predicted
line width (model) versus the actually observed one FWHM(observed).
Ideally, one would expect all the points to be on the diagonal.
However, the scatter in the parameters $\eta_x$ and $nU$ 
(which, in the analytic model we have assumed to have a single value for
all objects)
and  eventualy in the  spectral shape of the EUV, can easily explain
the scatter in the plot; infact, it is surprising the scatter is not
larger than a modest factor of ~3.

The  ratio between the observed FWHM and the calculated
velocity is scattered about unity, and shows no significant
dependence on the other observables ($L_x$ or $\alpha$), as is expected
if the model gives the correct dependence on these observables. 
The ratio appears to be weakly anticorrelated with the
observed FWHM (Fig. 3b), in the sense that very narrow line objects
give larger values for the predicted/observed FWHM ratio, and vice verca.
However, the correlation coeffient is $\rho=-0.2$ which is insignificant, 
certainly compared with the $\rho=0.7$ correlation coefficient of the
predicted vs. observed FWHM. This indicates that the model probably
takes the main effects into account, but there is a systematic residual
dependence on the line width. 

\subsection{Luminosity-dependent $L_x/M$}

What does this residual dependence tell us? 
If the basic assumptions made above are correct,
the bias has to be sought in the parameters. If, for example, $L_x/M$ 
depended on FWHM, so that it {\it decreased} with increasing FWHM, this
would reduce the effect, since for objects with larger FWHM Eq. (8) would 
give larger calculated values  and vice versa. 
One  explanation for such a dependence is
that objects with narrow lines tend to have a smaller L/M ratio, that is, 
narrow line objects tend to have relatively smaller mass black holes.

We now present a different, explanation, by showing 
that such a dependence does indeed follow from the well known phenomenological
relation between the optical-UV luminosity ($L_o$) and X-ray luminosity in AGN.
In transforming the black-hole mass to luminosity 
we have assumed that the Eddington ratio  ($\sim L/M$) 
is approximately the same for all
objects. Since we use the X-ray luminosity, we should actually take into account
the ratio $L_x/M$, or, $L_x/L$. 
It has been shown that in large samples of AGN 
$L_x\propto L_o^{0.75\pm0.05}$ (Kriss 1988, Mushotzky and 
Wandel 1989; 
note however that there is a considerable scatter around this relation).
If so, assuming $L_o/M\sim$ const. implies (on average)
$$\eta_x\propto L_x/M\propto L_o^{-0.25}\sim L_x^{-0.33}. \eqno (10)$$
We can plug this into the theoretical relation (Eq. (8) or (9)) to get
a phenomenological relation, which gives a modified dependence on 
$L_x$: 
$$FWHM\propto L_x^{0.25}\eta_x^{-1/2}\propto L_x^{0.4}. \eqno (11)$$

In order to see whether this explains the residual dependence on
FWHM (Fig. 3b), we should express the FWHM calculated/observed ratio 
in terms of FWHM instead of $L_x$.
Expressing $L_x$ in terms of FWHM in Eq. (8) and using  Eq. (10) gives 
$\eta_x\propto FWHM ^ {-0.8}$.
Since the values of FWHM calculated in Fig. 3b assume a fixed value for
$L_x/M$,  the calculated/observed FWHM ratio will have a bias of 
$1/\eta_x^{-1/2}\sim FWHM ^{-0.4}$
which agrees with the dependence seen in Fig. 3b.

\bigskip

\subsection{Luminosities}

%\begin{figure*}
%  \begin{minipage}{8.7cm}
%       \psfig{figure=final_fig4a.ps,height=8.7cm,width=8.7cm,clip=}
%\end{minipage}
%  \begin{minipage}{8.7cm}
%       \psfig{figure=final_fig4b.ps,height=8.7cm,width=8.7cm,clip=}
%\end{minipage}
%> Fig. 4 $
\begin{figure}
       \psfig{figure=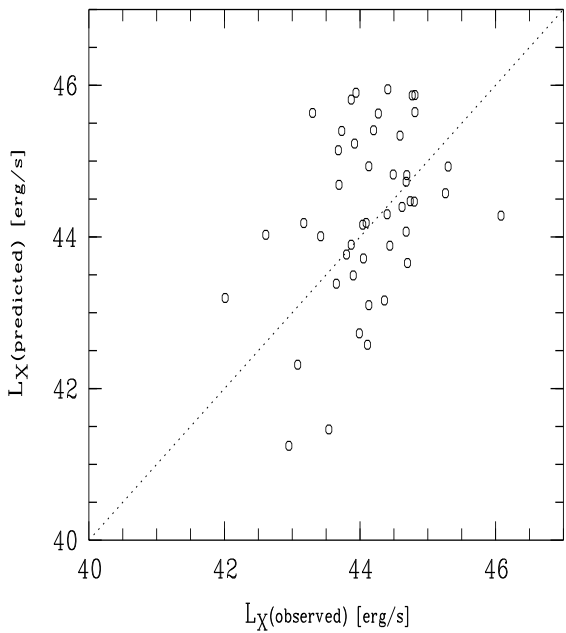,height=8.7cm,width=8.0cm,clip=}
\caption{
%Left panel (4a): Upper limit of the predicted ionizing luminosity vs. observed ROSAT
%luminosity. We assume an Eddington ratio of $\eta = 1$ to
%determine an upper limit for $L_{ion}$ following Eq. 6.
%The error bars correspond to the uncertainty in the energy index $\alpha$.
%Right panel (4b):
Luminosity prediction from the model.
Upper limit of the predicted ROSAT luminosity vs. observed ROSAT
luminosity. We assume an Eddington ratio of $\eta = 1$ and a relation between
ionizing and ROSAT luminosity of $L_{ion} = 7^{\alpha-1} L_x$.
%The error bars correspond to the uncertainty in the energy index $\alpha$.
}
\end{figure}

Eq. (8) predicts that the X-ray luminosity of individual objects
should increase in the direction normal to curves of constant
$L_x$ in the FWHM-$\alpha$ plane (see Fig. 1).
Fig. 2 gives  the actual observed ROSAT luminosities
(the size of the symbols corresponds to L) indicating a general increase of
the observed luminosities as predicted by the analytical relation.
Reversing Eq. (9)  and using the observed FWHM we have
$$L_{x45}(pred)=7^\alpha \left ({\alpha -1\over \alpha} 
\right ) E_0^{1-\alpha}$$
$$\left ( {FWHM\over 5600 km/s}\right )^4 
(n_{10}U)^{-1}(\eta_x/0.1)^2, \eqno (12)$$
so we may  calculate a {\it predicted} value for the X-ray (ROSAT band) 
luminosity, assuming e.g. $\eta_x =0.1$ (Fig. 4).
Comparing this with the observed ROSAT luminosity may be an alternative
test of the model. Theoretically, if there were no scatter in the parameters,
all objects should lie on the diagonal. The scatter in Fig. 4
appears larger
than in Fig. 3a, since the parameters are in the fourth power.
We also note that the slope of the best linear regression appears 
larger than unity, which reflects the residual dependence seen in Fig. 3b. 
If we assume $\eta_x$ does not depend on $L_x$, Eqs. (11) and (12) 
 give $L_x(pred)=L_x(obs)^{1.6}$,
 approximately the right slope for the regression in Fig. 4.

%Objects with observed ROSAT luminosities above the predicted upper limits
%corresponding to $\eta=1$
%(objects below the diagonal line in figs. 4a,b) would require Eddington ratios
%larger than 1 (or values of $Un>10^{10}$).

\section{Determination of M from variability analyses}

In Sect. 2 (Eq. 8) we have assumed all objects have the same
mass-luminosity ratio to derive an estimate for the black
hole mass. 
In order to test this assumption we make
below an alternative derivation with an independent estimate of the mass.

%> Figs 5 %
\begin{figure}
       \psfig{figure=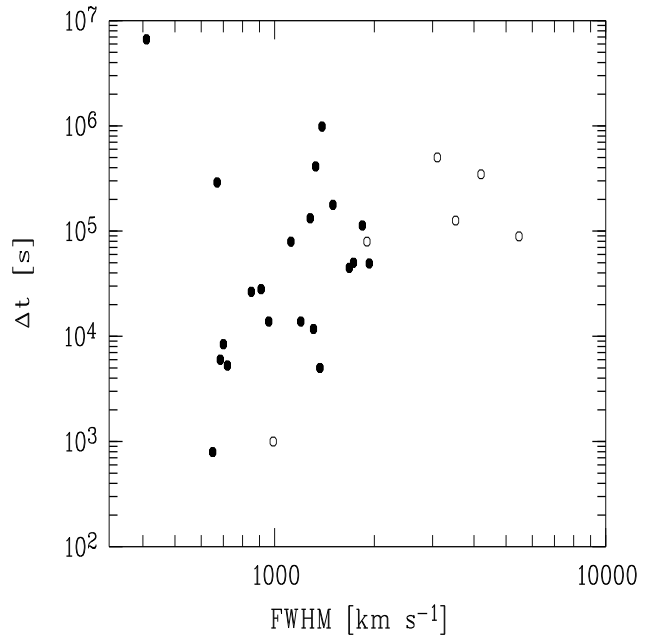,height=8.7cm,width=8.7cm,clip=}
\caption{Parameters determined from variability analyses.
Doubling time versus FWHM of $H\beta$ for NLS1 (filled circles)
and normal Seyfert 1s (open circles).
}
\end{figure}

\begin{figure*}
\begin{minipage}{8.7cm}
       \psfig{figure=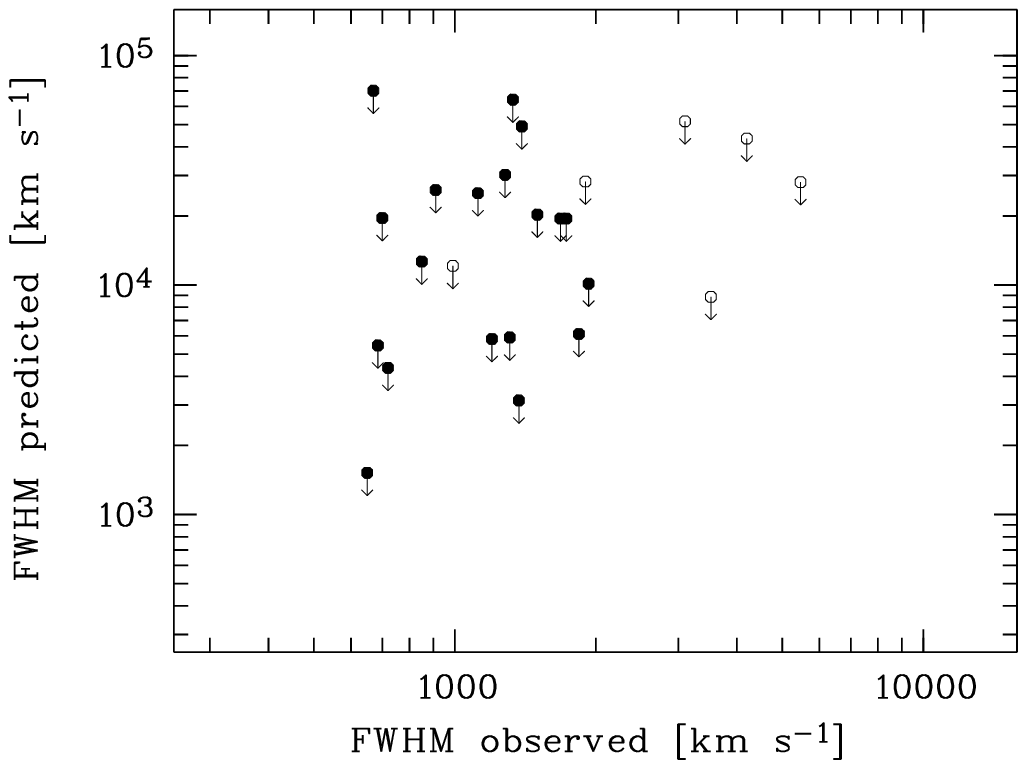,height=8.7cm,width=8.7cm,clip=}
\end{minipage}
\begin{minipage}{8.7cm}
       \psfig{figure=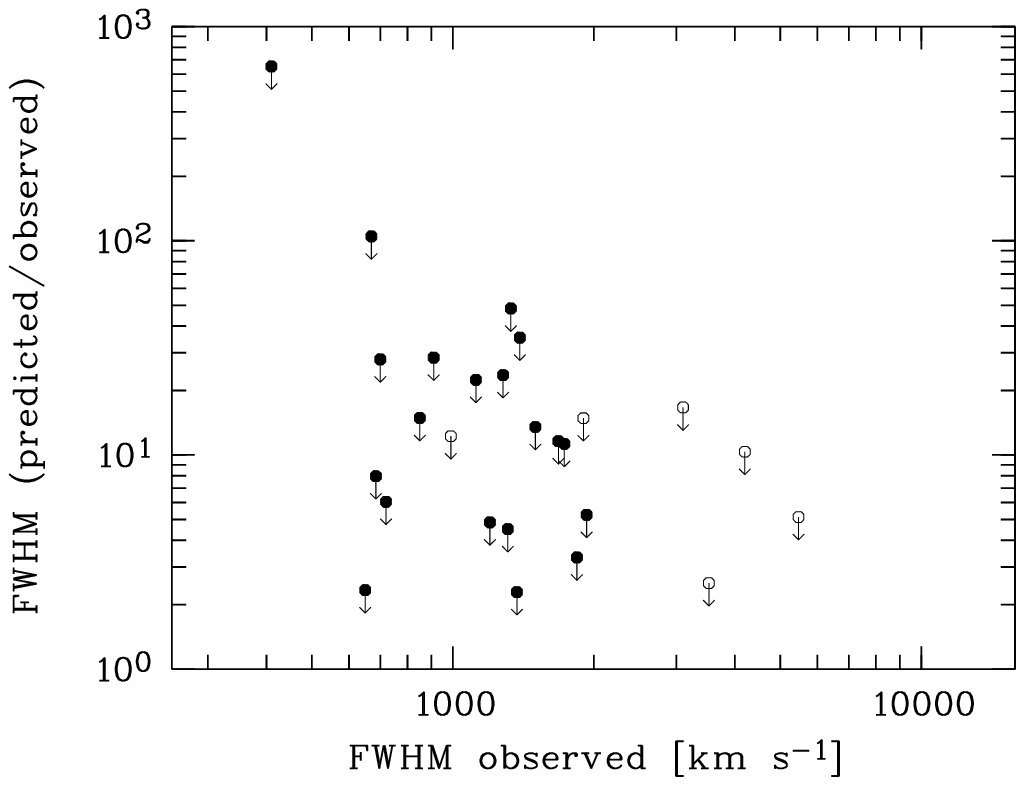,height=8.7cm,width=8.7cm,clip=}
\end{minipage}
\caption{FWHM predicted from the model using X-ray variability analyses
to set an upper limit on the black hole mass (Eq. (9)).
Left Panel (6a): Prediced vs. observed FWHM.  
Right pannel (6b): FWHM(observed)/FWHM(predicted)  vs. FWHM(observed) for
the objects with doubling times (Eq. 14)
}
\end{figure*}

An upper limit for the black hole mass is given by (e.g. Wandel \&
Mushotzky 1986)
$$M(\Delta t) < (c^3 / 10 G) \Delta t \simeq 10^4 \Delta t\ M_{\sun}, \eqno
(13)$$
where $\Delta t$ is in seconds.
This relation assumes that the bulk of the X-ray continuum is emitted within 5
Schwarzschild radii, and that the variability is not affected by beaming or relativistic
 motions (cf. Boller et al. 1997 for an example of relativistic boosting
 effects in a radio-quiet, ultrasoft NLS1).

Using the doubling time as the characteristic time for variability we
determine upper limits for the black hole mass for those objects
of BBF96 and Wandel \& Mushotzky (1986) whose continuum luminosities have been
observed to vary significantly.
Fig. 5 gives the doubling time (we have extrapolated amplitude
variations
linearely to a factor of 2 to determine $\Delta t$)  versus the observed
FWHM. Although there is some scatter, there is a strong indication for an
increase of the doubling time with observed FWHM.
This may indicate that NLS1 tend to have lower-mass black holes than
ordinary Seyfert 1 galaxies.

Finally, for the objects with an established doubling time and mass
upper limit, we may calculate the model velocity
without having to assume any $L/M$ relation.
Combining Eqs.  (7) and (12) we have
$$FWHM < (3500 km/s) \left(\frac{\Delta t}{10^4 s}\right )^{1/2}
\left ( \frac {\alpha}{\alpha-1}\right )^{1/4}L_{x45}^{-1/4} .
\eqno (14) $$

Fig. 6a shows the predicted FWHM upper limits,
and 6b - the ratio FWHM(observed)/FWHM(predicted),  vs.
FWHM(observed) for the objects with doubling times.
We see that the predicted FWHM is larger than the observed
one by an average factor of 10, which is not surprising, considering
that the variability method gives only an upper limit for the mass.
Similarly to the photoionization method, also here the ratio shows
a slight anticorrelation with the observed FWHM.

\section{Summary and Discussion}
\subsection{Basic Model}
We derive a simple theoretical model which explains the observed distribution
of line width and spectral index for Seyfert 1 galaxies.
%The basic ideas are as follows:
The observed velocity dispersion is determined by the mass of the
black hole and the broad line region radius.
Sources with steep X-ray spectra (for a given luminosity)
have a stronger ionizing power than flat-spectrum sources, which implies
that the BLR is formed at relatively larger distances from
the central source, and hence has a smaller velocity dispersion
and a smaller observed FWHM.

\subsection{Model Parameters}
The model predicts a particular distribution of the luminosities
in the three dimensional FWHM-$L_x-\alpha$ space, which is essentially confirmed by the data.
The FWHM also depends on the parameters $U, n, E_0$
(and perhaps other parameters we are not aware of), as shown in Eqs. (8)
and (9). All those parameters, however, cannot be observed directly or are
not available
for the individual objects, so we develop and test the explicit teoretical
relation for the observables that are available for the individual objects
in our sample, and show that the data is consistent with the theoretical
relation we derive. The dependence on the other parameters is kept in
the equations, but because of our observational ignorance we assume
that all objects have the same values for those parameters. This choise
leads to an implicit error, which contributes to the
scatter in Fig. 3. Because of the weak dependence on the parameters, the
expected error is relatively small, of the order of 3 for the
expected parameter range. 

In addition to the scatter in the parameters, there may be also a systematic
dependence on some of the observables, which would lead to a dependence
of the predicted/observed FWHM ratio. 
This is shown to be the case for the parameter
$\eta_x=L_x/L_{Edd}$, and is treated in Sect. 3.2. In principle, also the other
 parameters may systematically depend on some variable, but the data required
to check this (e.g. separate determinations of $Un_e$) for individual objects
are not yet available.

\subsection{Observational Bias}
 BBF96 searched the literature for reports of NLS1 and found 46 in total.
 32 of these NLS1 were located in in the fields of view of ROSAT pointings
 in the public archive. Only one of these 32 objects was not found above a
 5 $\sigma$ detection limit.
 Walter \& Fink 1993 selected all Seyfert 1 type AGN with more than
 300 counts in the ROSAT All-Sky Survey data which were observed at least
 once in the ultraviolet with IUE. Although the selection of broad and
 narrow-lined AGN is based on these special selection criteria,
 there are no obviousg selection effects that can  account
 for the absence of broad line AGN with steep soft X-ray spectra. 
The measurements in the photon index and the H$_{\beta}$
 values are totally independent so that there is no obvious way a
 spurious correlation could be induced.

\subsection{Independent determination of the Mass}
Estimating an upper limit for the
central mass directly from X-ray variability data we find that AGN with
narrow optical emission lines may have lower black hole masses than
broad line AGNs (Fig. 5). In a separate work we will consider physical
parameters leading to soft X-ray spectra to further improve our model.
In order to test this mass estimate, it is possible to compare the BLR
radius calculated from the ionizing luminosity (e.g. Eq. 2) or from
reverberation mapping to the radius inferred from the observed FWHM and
the mass estimated from the variability. Doing this indicates that the
variability mass tends to be larger than the real mass for low narrow-line
AGN (Wandel and Boller 1997).

\subsection{What causes steep X-ray spectra in NLS1s?}
We have shown that sources with a steep soft X-ray spectrum will have
narrower broad lines than objects with flat spectra.
A  question we have not touched in this work is the cause for the observed 
steep X-ray spectra in many NLS1s. In principle, smaller central masses 
could produce  
steep soft X-ray spectra as follows.
For a given accretion rate, lower mass black holes could yield a hotter
accretion disk,which
would radiate more energy in the soft X-ray band. 
X-ray reprocessing by the accretion disk would produce a
soft X-ray spectrum (Matsuoka et al. 1990, Pounds et al. 1990) that extends
 to higher energies. This  would make them appear to have steeper ROSAT 
spectra since there would be more flux in the soft X-ray band. 
This scenario will be considered in a separate work.

{}

\vskip 1.0cm
\acknowledgements
We thank the referee, T. Courvoisier for valuable remarks and
 Hagai Netzer for useful discussions.
AW Thanks Max-Planck-Institut f\"ur extraterrestrische Physik
for the kind hospitality.

\vskip 2.0cm
\begin{thebibliography}{}

\bibitem[]{} Alexander, T., Netzer, H., 1994, MNRAS, 270, 781

\bibitem[]{} Boller, Th., Brandt, W.N., Fink, H. 1996,  A\&A, 305, 53 (BBF96)

\bibitem[]{} Boller, Th., Brandt, W.N., Fabian, A.C., Fink, H., MNRAS submitted; astro-ph/9703126. 

\bibitem[]{} Clavel J. et al. 1991, ApJ, 366, 64.

\bibitem[]{} Kriss, G.A. 1988, Apj , 324, 809.

\bibitem[]{} Matsuoka  et al. 1990, ApJ, 361, 440.

\bibitem[]{} Mushotzky, R.F. and Wandel, A. 1989, Apj , 339, 674.

\bibitem[]{} Netzer, H. 1990 in "Active Galactic Nuclei", Saas-Fee Advanced Course
20, eds. T.J.-L. Courvoisier and M. Major, Springer Verlag, Berlin, p. 57.

\bibitem[]{} Peterson, B.M. in "Reverberation Mapping of AGN" eds. P.M.
Gondhalekar, K.Horne,  B.M.Peterson, SFASP 1995.

\bibitem[]{}Rees, M., Netzer H., Ferland, G.J. 1989. ApJ , 347, 640.

\bibitem[]{} Pounds, K. et al., 1990, Nature, 344, 132.

\bibitem[]{}Walter, R. and Fink, H.H. 1993, A\&A , 274, 105.

\bibitem[]{}Wandel, A. 1996, in "X-ray Imaging and Spectroscopy of Cosmic
Hot Plasmas", p. 307, ed. F.Makino, K. Mitsuda, Universal Academy Press, Tokyo .

\bibitem[]{}Wandel, A. 1997, ApJ Letters, 490, in press.

\bibitem[]{}Wandel, A. and Mushotzky, R.F. 1986, ApJ  , 306, L61.

\bibitem[]{}Wandel, A. and Yahil, A. 1985, ApJ  , 295, L1.

\bibitem[]{}Wandel, A. and Boller, Th. 1997, in "Astronomical Time Series",
 eds. D. Maoz et.al., Dordrecht: Kluwer, p. 255; astro-ph/9703198. 


\end{thebibliography}
\end{document}